# Local Pauli stabilizers of symmetric hypergraph states

David W. Lyons, Nathaniel P. Gibbons, Mark A. Peters,
Daniel J. Upchurch, Scott N. Walck, and Ezekiel W. Wertz

revised: 10 April 2017

## Abstract

Hypergraph states of many quantum bits share the rich interplay between simple combinatorial description and nontrivial entanglement properties enjoyed by the graph states that they generalize. In this paper, we consider hypergraph states that are also permutationally invariant. We characterize the states in this class that have nontrivial local Pauli stabilizers and give applications to nonlocality and error correction.

## 1 Introduction

Entangled states of many quantum bits are essential resources in emerging technologies including ultra-secure communication, powerful computing machines, and measuring devices that promise to outperform 'classical' digital technologies in fundamental ways. While such applications drive the study of multiparticle entanglement, a deeper motivation is to better understand quantum mechanics and foundations of physics.

In its full generality, entanglement is a hard problem: it is not reasonable to expect the achievement of a full classification of multiparticle entanglement types [1]. Nevertheless, studies of entanglement have illuminated operational questions that facilitate applications and theoretical understanding. An example is the class of graph states, which admit a concise combinatorial description and are important not only for applications including measurement-based quantum computation and error-correcting codes, but also for theoretical insight on entanglement through the Pauli stabilizer formalism.

Recent papers [2, 3, 4, 5, 6, 7, 8, 9, 10, 11, 12, 13] have investigated *hypergraph* states that share the rich interplay between simple description and nontrivial entanglement properties enjoyed by the graph states that they generalize. Introductions in references [6, 8, 10] provide surveys and references for applications of hypergraphs in quantum algorithms and demonstrations of nonlocality. These same papers have established basic results on entanglement properties of hypergraph states. Gühne et al. [6] identify hypergraph states as a subclass of the locally maximally entangled (LME) states, classify local unitary inequivalent hypergraph states up to 4 qubits, and give general conditions for local unitary and local Pauli equivalence of generic hypergraph states. In our paper [10] we give conditions for the existence of continuous local unitary stabilizer subgroups and identify infinite families of hypergraph states with discrete local Pauli stabilizers. In [9], Tsimakuridze and Gühne use hypergraph states to construct infinite families of graph states that are interconvertible by local unitary (LU) operations, but not by local Clifford (LC) operations, thereby providing new insight on the failure of the so-called LU-LC conjecture for graph states; their hypergraph methodology provides some structural understanding of previously known LU-LC counterexamples, first found by Ji et al. [14] using ad hoc methods. Papers [11, 12, 13] provide further analysis of LU operations on hypergraph states, extend hypergraph constructions to qudits, and develop a verification protocol for measurement-based quantum computation using hypergraph states. This brief survey of theory and applications underscores the richness of hypergraph states for analyzing and harnessing multiparticle entanglement.

In this paper we consider symmetric (that is, permutationally invariant) hypergraph states. Our main theoretical result is a classification of symmetric hypergraph states that have nontrivial local Pauli stabilizers. We illustrate some applications that demonstrate why these states are interesting: states that have a certain local Pauli stabilizing operator can be used to construct violations of noncontextuality inequalities (this extends a result in [6]); and any symmetric hypergraph state with nontrivial local Pauli stabilizer can be used to construct a code that corrects so-called collective decoherence errors. Finally, we show how symmetric hypergraph states are robust to the loss of one or more qubits (while this is not dependent on any local stabilizer, it is an interesting addendum to the error correction result).

The paper is organized as follows. Because there are many statements whose proofs are long and technical, we segregate the proofs into the appendix. In Section 2 we establish notation and define basic objects. Statements of the main theoretical Theorems and Lemmas are in Section 3. This is followed by an application to nonlocality (Section 4); a code construction with error correction (Section 5); and reconstruction from reduced density matrices of subsystems (Section 6). Section 7 concludes with remarks on future directions.

## 2 Preliminaries

### Hypergraphs and hypergraph states

A hypergraph is a pair $G = (V, E)$ where $V$ is a set of *vertices* and $E$ is a collection of nonempty subsets of $V$ called the *hyperedges* of $G$. Ordinary graphs (undirected, with no loops and no parallel edges) are precisely those hypergraphs whose hyperedges all have cardinality 2. We shall write $|e|$ to denote the cardinality of the hyperedge $e$. [Note: Some authors allow the possibility of an empty hyperedge. In this paper the cardinality of any hyperedge is a positive integer.]

Given a hypergraph $G$ with $n$ vertices labeled $1, 2, \ldots, n$, we define the corresponding $n$-qubit quantum hypergraph state $|G\rangle$ by

$$|G\rangle := \left( \prod_{e \in E} C_e \right) |+\rangle^{\otimes n} \qquad (1)$$

where $|+\rangle$ denotes the 1-qubit "plus" state $|+\rangle = \frac{1}{\sqrt{2}} (|0\rangle + |1\rangle)$ and $C_e$ denotes the *generalized controlled-Z operator*, defined by its action on the computational basis vector $|I\rangle = |i_1 i_2 \cdots i_n\rangle$ (each $i_k = 0, 1$) as follows.

$$C_e |I\rangle := (-1)^{\prod_{k \in e} i_k} |I\rangle \tag{2}$$

Said another way, $C_e$ acts on the subsystem of qubits in $e$ by

$$|J\rangle \to \begin{cases} -|J\rangle & \text{if } j_1 = j_2 = \cdots = j_{|e|} = 1 \\ |J\rangle & \text{otherwise} \end{cases}$$

where $J = j_1 j_2 \ldots j_{|e|}$ is a bit string of length $|e|$, and $C_e$ acts as the identity on remaining qubits in the complement of $e$. Note that when $|e| = 2$, then $C_e$ is precisely the controlled-$Z$ operator on the 2 qubits in $e$ (hence the name "generalized controlled-$Z$"). We say a hypergraph or its corresponding state is $m$-uniform if all hyperedges contain exactly $m$ vertices. In this language, a graph is a 2-uniform hypergraph and a graph state is a hypergraph state for which all of the $C_e$ operators are ordinary controlled-$Z$ gates. Finally, regardless of whether or not the empty hyperedge is allowed, it is convenient to define operator $C_\emptyset$ to be the negative of the identity operator.

## Symmetric hypergraph states, $|K_n^m\rangle$ notation

An $n$-qubit state is said to be symmetric, or permutationally invariant, if it is unchanged by permutations of the qubits. For example, $|00\rangle + |01\rangle$ is not symmetric because transposing qubits 1 and 2 produces the state $|00\rangle + |10\rangle \neq |00\rangle + |01\rangle$, while $|01\rangle + |10\rangle$ is symmetric. One readily sees that a hypergraph state is symmetric if and only if, for each hyperedge $e$, the set $E$ of hyperedges contains all possible hyperedges of cardinality $|e|$. We say a hypergraph is $m$-complete if it contains all possible hyperedges of cardinality $m$. Thus the symmetric hypergraphs are precisely those which are complete for some list of positive integers $m_1, m_2, \ldots, m_k$.

We generalize the graph theoretic notion of the $n$-vertex complete graph, that is, the graph on $n$ vertices that has all possible $\binom{n}{2}$ edges, standardly denoted $K_n$. We define the $n$-vertex, $(m_1, m_2, \ldots, m_k)$-complete hypergraph, denoted $K_n^{m_1, m_2, \ldots, m_k}$, to be the $n$-vertex hypergraph that is complete in cardinalities $m_1, m_2, \ldots, m_k$ and contains no hyperedges of any other cardinalities. For readability, we consolidate the hyperedge cardinality vector with a single symbol $m = (m_1, m_2, \ldots, m_k)$, and write $K_n^m$ to denote $K_n^{m_1, m_2, \ldots, m_k}$. Throughout, we shall always assume that the integers $n, m_1, m_2, \ldots, m_k$ satisfy $1 \leq m_1 < m_2 < \cdots < m_k \leq n$. Given integers $m, n$ satisfying $1 \leq m \leq n$, we will write $K_n^m$ to denote the $n$-qubit symmetric hypergraph which is complete in the single hyperedge cardinality $m$. Whether $m$ denotes a vector of integers or a single integer in the symbols '$K_n^m$' will be clear from context.

In this notation, the complete graph $K_n$ is the 2-complete hypergraph $K_n^2$. We write $|K_n^m\rangle$ (or $|K_n^{m_1, m_2, \ldots, m_k}\rangle$) to denote the hypergraph state with hypergraph $K_n^m$ (or $K_n^{m_1, m_2, \ldots, m_k}$). By the remarks in this subsection, we can say that if $|G\rangle$ is an $n$-qubit symmetric hypergraph state, then it is of the form $|K_n^m\rangle$ for some list of positive integers $m = (m_1, \ldots, m_k)$. Figure 1 illustrates the definitions in this subsection.

### Computational basis state vector coefficients

Any $n$-qubit pure symmetric state $|\psi\rangle$ (whether or not it is a hypergraph state) expressed in the computational basis

$$|\psi\rangle = \sum_I c_I |I\rangle$$

must satisfy $c_I = c_{I'}$ whenever $\mathrm{wt}(I) = \mathrm{wt}(I')$, where $\mathrm{wt}(I) := \sum_j i_j$ denotes the Hamming weight of the bit string $I = i_1 i_2 \ldots i_n$ (each $i_j = 0, 1$). Thus we can gather the terms by weight, with a single common coefficient $d_w$ for all weight $w$ standard basis vectors in the expansion of $|\psi\rangle$.

$$|\psi\rangle = \sum_{w=0}^n d_w \left( \sum_{I:\, \mathrm{wt}(I)=w} |I\rangle \right)$$

In the case of a symmetric hypergraph state $|G\rangle$, we define coefficients $f_w = \pm 1$ by

$$\sqrt{2^n}\, |G\rangle = \sum_{w=0}^n f_w \left( \sum_{I:\, \mathrm{wt}(I)=w} |I\rangle \right) \tag{3}$$

and we will refer to $(f_0, f_1, \ldots, f_n)$ as the 'weight sign vector' of $|G\rangle$.

### Operations on hyperedge cardinality vectors

Statements and arguments in this paper are frequently recursive in nature, with comparisons between states of the form

$$|K_n^{m_1,\ldots,m_k}\rangle,\, \left|K_{n-1}^{m_1-1,\ldots,m_k-1}\right\rangle,\, \left|K_{n+1}^{m_1+1,\ldots,m_k+1}\right\rangle.$$

For convenience and readability, we will write $\left|K_{n-1}^{m-1}\right\rangle, \left|K_{n+1}^{m+1}\right\rangle$ to denote the states $\left|K_{n-1}^{m_1-1,\ldots,m_k-1}\right\rangle, \left|K_{n+1}^{m_1+1,\ldots,m_k+1}\right\rangle$, obtained by adding or subtracting 1, respectively, to all of the integers $n, m_1, \ldots, m_k$. In the case that $m_1 = 1$, we simply drop the leading zero entry from the list $m - 1$, so that

$$m - 1 = (m_2 - 1, \ldots, m_k - 1),$$

(we do not consider hypergraphs having an empty hyperedge). Lastly, we write $\left|K_n^{1,m}\right\rangle$ to denote the state $\left|K_n^{1,m_1,\ldots,m_k}\right\rangle$ obtained by prepending the entry 1 in the hyperedge cardinality vector.

## 3 Local Pauli stabilizers: statements of results

Our goal is to characterize symmetric hypergraph states that are stabilized by local Pauli operations. We use the phrase "$|\psi\rangle$ is $U_1 \otimes U_2 \otimes \cdots \otimes U_n$-stable" to mean that $(U_1 \otimes U_2 \otimes \cdots \otimes U_n) |\psi\rangle = |\psi\rangle$ for an $n$-fold tensor product of local unitary operators $U_k$ operating on an $n$-qubit pure state $|\psi\rangle$. In this section, we show that the problem of our goal reduces to the characterization

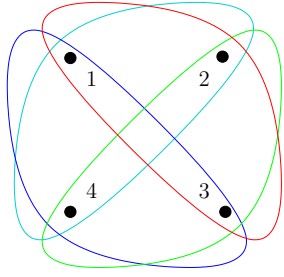

$$\begin{aligned}
\left|K_4^3\right\rangle &= C_{123}C_{234}C_{341}C_{412}\left|+\right\rangle^{\otimes 4}\\
&= \left|0000\right\rangle\\
&+ \left|1000\right\rangle + \left|0100\right\rangle + \left|0010\right\rangle + \left|0001\right\rangle\\
&+ \left|1100\right\rangle + \left|1010\right\rangle + \left|1001\right\rangle + \left|0110\right\rangle + \left|0101\right\rangle + \left|0011\right\rangle\\
&- \left|1110\right\rangle - \left|0111\right\rangle - \left|1011\right\rangle - \left|1101\right\rangle\\
&+ \left|1111\right\rangle
\end{aligned}$$

Figure 1: The symmetric hypergraph $K_4^3$ and its corresponding state. Colors show the matching between hyperedges $e$ and the weight 3 basis vectors on which their corresponding $C_e$ gates act. All 4 gates act on the last term $\left|1111\right\rangle$ with the net sign effect $(-1)^4 = 1$.

of symmetric hypergraph states that are either $X^{\otimes n}$-stable, $-X^{\otimes n}$-stable, or $Y^{\otimes n}$-stable, where $X = \begin{bmatrix} 0 & 1 \\ 1 & 0 \end{bmatrix}$ and $Y = \begin{bmatrix} 0 & -i \\ i & 0 \end{bmatrix}$ are the Pauli matrices.

We begin with a Lemma that expresses the relationship between the hyperedge cardinality vector $m = (m_1, m_2, \ldots, m_k)$ and the weight sign vector (defined by (3)) that encodes the state vector coefficients of the symmetric hypergraph state $\left|K_n^m\right\rangle$ in the computational basis. The key ingredient is Pascal's triangle mod 2. See Figure 2 for an illustration of the objects in Lemmas 1 and 2.

**Lemma 1** *Let integers $n, m_1, m_2, \ldots, m_k$ satisfy $1 \leq m_1 < m_2 < \cdots < m_k \leq n$, and consider the symmetric hypergraph state $\left|K_n^m\right\rangle$ where $m$ is the hypergraph edge cardinality vector $m = (m_1, \ldots, m_k)$, with weight sign vector $(f_0, f_1, \ldots, f_n)$. Let $e = (e_0, \ldots, e_n)$ denote the vector with $0, 1$ entries such that*

$$f_w = (-1)^{e_w}.$$

*Let $g = (g_0, \ldots, g_n)$ be the vector with $0, 1$ entries given by*

$$g_w = \begin{cases} 1 & \text{if } w = m_k \text{ for some } k \\ 0 & \text{otherwise} \end{cases}.$$

*Let $A$ be the $(n+1) \times (n+1)$ matrix whose $i, j$ entry is $\binom{i}{j}$ (mod 2), for $0 \leq i, j \leq n$ (with the usual convention that $\binom{i}{j} = 0$ for $i < j$). We have the following equations mod 2.*

$$\begin{aligned}
e &= Ag\\
A &= A^{-1}\\
g &= Ae
\end{aligned}$$

The following Lemma characterizes symmetric hypergraph states with nontrivial local Pauli symmetry in terms of conditions on sums of entries in Pascal's triangle mod 2. We call them "palindrome conditions" because (4) literally says that a the weight sign vector (3) associated to the hypergraph state is unchanged if your read its entries in reverse order. Conditions (5) and (6) say that modifications of the weight sign vector vector are palindromes. Details are in the appendix.

**Lemma 2 (Palindrome conditions**[1]**)** *Let $n$ be a positive integer and let $m = (m_1, m_2, \ldots, m_k)$ be a vector of positive integers satisfying $1 \leq m_1 < m_2 < \cdots < m_k \leq n$. The symmetric hypergraph state $|K_n^m\rangle$ is:*

(i) $X^{\otimes n}$-*stable if and only if*

$$\sum_{j=1}^{k} \binom{w}{m_j} = \sum_{j=1}^{k} \binom{n-w}{m_j} \pmod{2}, \quad 0 \leq w \leq n \qquad (4)$$

(ii) $-X^{\otimes n}$-*stable if and only if*

$$\sum_{j=1}^{k} \binom{w}{m_j} = \sum_{j=1}^{k} \binom{n-w}{m_j} + 1 \pmod{2}, \quad 0 \leq w \leq n \qquad (5)$$

(iii) $Y^{\otimes n}$-*stable if and only if $n$ is even and*

$$\sum_{j=1}^{k} \binom{w}{m_j} = \sum_{j=1}^{k} \binom{n-w}{m_j} + w + n/2 \pmod{2}, \quad 0 \leq w \leq n \qquad (6)$$

The next result says that symmetric hypergraph states impose restrictions on their local unitary stabilizing operators: except for graph states (2-uniform hypergraph states), any local unitary operator that stabilizes a symmetric hypergraph state with 3 or more qubits must act by the *same* unitary in every qubit. Here is the formal statement.

**Theorem 1** *Let $|G\rangle$ be an n-qubit symmetric hypergraph state with $n \geq 3$, is not a graph state (that is, $G$ is not of the form $K_n^2$), and suppose that $\bigotimes_j U_j = U_1 \otimes U_2 \otimes \cdots \otimes \cdots U_n$ is a local unitary stabilizing operator for $|G\rangle$. Then there exists an operator $U \in U(2)$ (which can be taken to be a phase times any one of the $U_j$) such that $\bigotimes_j U_j = e^{it} U^{\otimes n}$ for some phase $e^{it}$.*

The next result is that local Pauli stabilizing operators for symmetric hypergraph states are limited to only three possibilities. A consequence is that the three cases in the palindrome condition Lemma 2 are in fact comprehensive.

**Theorem 2** *If a symmetric hypergraph state of 3 or more qubits (that is not a graph state) has a nontrivial local Pauli stabilizing operator, then that stabilizing operator is of the form $X^{\otimes n}$, $-X^{\otimes n}$, or $Y^{\otimes n}$.*

---
[1] This Lemma generalizes a special case given in our earlier paper [10] and also corrects a minor error. See the proof in the appendix for details.

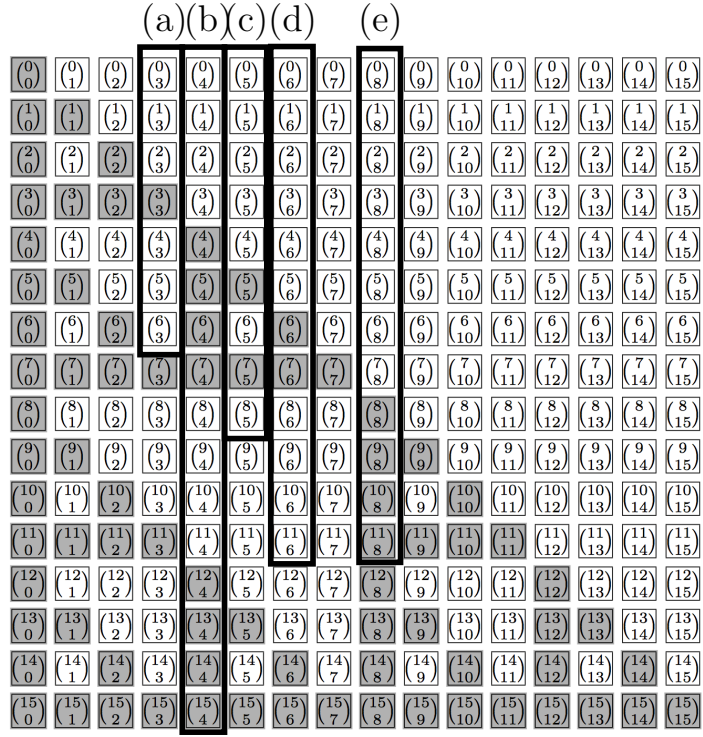

Figure 2: Pascal's triangle mod 2, sign weight vectors, and palindrome conditions for local Pauli stabilizers. Colors gray and white indicate parity of entries in the matrix of binomial coefficients: gray is odd and white is even. Reading each white entry as 0 and each gray as 1, Pascal's triangle mod 2 appears below the main diagonal of the matrix. In the notation of Lemma 1, the illustration shows the matrix $A$ for $n = 16$. One reads directly the sign weight vector for $|K_n^m\rangle$ in the $\binom{\cdot}{m}$ column: the sign weight vector entry $f_w$ is 1 if $\binom{w}{m}$ is white and is $-1$ if $\binom{w}{m}$ is gray. Column vector (a) is the palindrome sign weight vector $(1, 1, 1, -1, 1, 1, 1)$ for the $X^{\otimes n}$-stable state $|K_6^3\rangle$. Vector (b) is the antipalindrome sign weight vector for the $-X^{\otimes n}$-stable state $|K_{15}^4\rangle$, and (c) is the weighted palindrome for the $Y^{\otimes n}$-stable state $|K_8^5\rangle$. The sum (d) + (e) is the antipalindrome for the $-X^{\otimes n}$-stable state $\left|K_{11}^{6,8}\right\rangle$.

| $\lvert K_n^m\rangle$ stability | $\Rightarrow$ | $\lvert K_{n-1}^{m-1}\rangle$ stability |
|---|---|---|
| $X^{\otimes n}$ | | $X^{\otimes n}$ |
| $-X^{\otimes n}$ | | $X^{\otimes n}$ |
| $Y^{\otimes n}$ | | $-X^{\otimes n}$ |

| $\lvert K_n^m\rangle$ stability | $\Rightarrow$ | $\lvert K_{n+1}^{m+1}\rangle$ stability, | $\lvert K_{n+1}^{1,m+1}\rangle$ stability |
|---|---|---|---|
| $X^{\otimes n}$ and $n$ odd | | $X^{\otimes n}$ | $X^{\otimes n}$ |
| $X^{\otimes n}$ and $n$ even | | $\pm X^{\otimes n}$ | $\mp X^{\otimes n}$ |
| $-X^{\otimes n}$ | | $Y^{\otimes n}$ | (none) |

Table 1: Going Down and Going Up Lemmas: Summary

The next two lemmas are the main facts that capture all the possibilities for what happens to local Pauli stabilizers when one adds or deletes a qubit. The statements are summarized in chart form in Table 1.

**Lemma 3 ("Going down")** *Suppose that the symmetric hypergraph state $\lvert K_n^m\rangle$ is (a) $X^{\otimes n}$-stable, (b) $-X^{\otimes n}$-stable, or (c) $Y^{\otimes n}$-stable. Then $\lvert K_{n-1}^{m-1}\rangle$ is (a) also $X^{\otimes n}$-stable, (b) $X^{\otimes n}$-stable, or (c) $-X^{\otimes n}$-stable, respectively.*

**Lemma 4 ("Going up")** *Suppose that the symmetric hypergraph state $\lvert K_n^m\rangle$ is $X^{\otimes n}$-stable. If $n$ is odd, then both $\lvert K_{n+1}^{m+1}\rangle$ and $\lvert K_{n+1}^{1,m+1}\rangle$ are $X^{\otimes n}$-stable. If $n$ is even, then exactly one of the states $\lvert K_{n+1}^{m+1}\rangle, \lvert K_{n+1}^{1,m+1}\rangle$ is $X^{\otimes n}$-stable, and the other is $-X^{\otimes n}$-stable. Now suppose that $\lvert K_n^m\rangle$ is $-X^{\otimes n}$-stable. Then $\lvert K_{n+1}^{m+1}\rangle$ is $Y^{\otimes n}$-stable.*

Finally, Theorems 3 and 4 give base cases and the inductive step that characterizes symmetric hypergraph states with local Pauli symmetry. Figure 3 illustrates the patterns in the base case.

**Theorem 3 (Symmetric hypergraph states with nontrivial local Pauli stabilizer: base case[2])** *Let $\lvert K_n^m\rangle$ denote a symmetric hypergraph whose hyperedge cardinality vector has only a single entry $m$, with $1 \leq m \leq n$.*

(i) *$\lvert K_n^m\rangle$ is $X^{\otimes n}$-stable if and only if $m$ is in the range $2^t \leq m \leq 2^{t+1}-1$ and $n = (s+1)2^{t+1} + m - 1$ for some $t \geq 0$, $s \geq 0$.*

(ii) *$\lvert K_n^m\rangle$ is $-X^{\otimes n}$-stable if and only if $m = 2^t$ and $n = (2s+1)2^t + 2^t - 1$ for some $t \geq 0$, $s \geq 0$.*

(iii) *$\lvert K_n^m\rangle$ is $Y^{\otimes n}$-stable if and only if $m = 2^t + 1$ and $n = s2^{t+1}$ for some $t \geq 0$, $s \geq 1$.*

---

[2]The 'if' direction for statement (i) of Theorem 3 is stated and proved in our previous paper [10]. The 'if' direction for statements (ii) and (iii) in this paper increases the range of the parameter $t$ to $t \geq 0$, rather than the range $t \geq 1$ given in [10]. Statements in both papers are correct: the version in this paper is more complete.

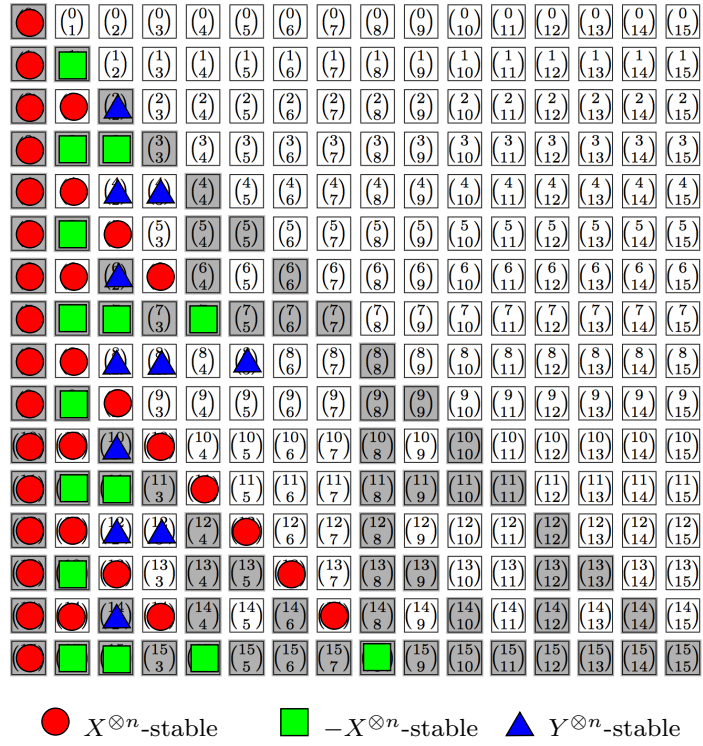

Figure 3: Theorem 3 illustration. $|K_n^m\rangle$ has local Pauli stabilizer indicated by the shape symbol placed on the entry $\binom{n}{m}$ in the grid.

**Theorem 4 (Symmetric hypergraph states with nontrivial local Pauli stabilizer: general case)** *Consider the symmetric hypergraph state $|K_n^m\rangle$ with hyperedge cardinality vector $m = (m_1, m_2, \ldots, m_k)$ satisfying $1 \leq m_1 < m_2 < \ldots < m_k \leq n$.*

*(a) $|K_n^m\rangle$ is $X^{\otimes n}$-stable if and only if $|K_{n-1}^{m-1}\rangle$ is $X^{\otimes n}$-stable, and*

$$\binom{n}{m_1} + \binom{n}{m_2} + \cdots + \binom{n}{m_k} = 0 \pmod{2}.$$

*(b) $|K_n^m\rangle$ is $-X^{\otimes n}$-stable if and only if $|K_{n-1}^{m-1}\rangle$ is $X^{\otimes n}$-stable and*

$$\binom{n}{m_1} + \binom{n}{m_2} + \cdots + \binom{n}{m_k} = 1 \pmod{2}.$$

*(c) $|K_n^m\rangle$ is $Y^{\otimes n}$-stable if and only if $|K_{n-1}^{m-1}\rangle$ is $-X^{\otimes n}$-stable.*

### Some consequences and comments

Theorems 3 and 4 imply a recursive formula and indeed a closed-form expression for the number of $n$-qubit $X^{\otimes n}$-stable symmetric hypergraph states (with smallest hyperedge cardinality at least 1). If we call this number $C_n$, we have

$$C_n = 2^{\lfloor n/2 \rfloor} - 1.$$

This is easy to see, starting with the base case $C_2 = 1$. Inductively, if $n$ is odd, then every hyperedge cardinality vector $m$ gives rise to two $n+1$ qubit hyperedge cardinality vectors, namely $m+1$ and $1, m+1$. There is also the "new" $n+1$ hyperedge cardinality vector 1 (that is, a vector with a single entry 1). So $C_n = r$ for $n$ odd implies $C_{n+1} = 2r + 1$. Similarly, going up from $n$ even to $n+1$ odd, we just promote each hyperedge cardinality vector $m$ to $m+1$. So the sequence $C_n$ ($n \geq 2$) goes

$$1, 1, 3, 3, 7, 7, 15, 15, \ldots.$$

Another observation, coming immediately from these results: If an $n$-qubit symmetric hypergraph state is $X^{\otimes n}$-stable, then it cannot have the hyperedge of cardinality $n$. Why? Because going down would take you to $|K_1^1\rangle$, which is not $X^{\otimes n}$-stable.

## 4 Application: nonlocality

In [6], Gühne et al. show that symmetric hypergraph states of the form $|K_n^m\rangle$ with $n, m$ meeting the conditions of Theorem 3 part (ii) can be used to demonstrate nonlocality through the violation of a hypergraph generalization of the Mermin noncontextuality inequality, originally for graph states. The property that makes this construction work is stability under the action of $-X^{\otimes n}$. In this section we extend this result to a wider class of $-X^{\otimes n}$-stable symmetric hypergraph states that have more than one completeness level (that is, that have hyperedges of more than one cardinality). In order to state the results, we give a brief recap of how the Mermin violation construction works.

Let $|G\rangle$ be an $n$-qubit symmetric hypergraph state. For qubit index $j$, let $E(j)$ denote the set of hyperedges containing $j$, and let $g_j = X_j \prod_{e \in E(j)} C_{e\setminus j}$ (here and following we write $C_{e\setminus j}$ in lieu of the more precise but cumbersome $C_{e\setminus \{j\}}$). Let $M$ denote the operator $M = \sum_{j=1}^n g_j + \prod_{j=1}^n g_j$. The hypergraph $G$ can be chosen so that the value of measurement $M$ will be $n+1$, while at the same time any assignment of values $\pm 1$ to classical variables $X_j, C_{e\setminus j}$ leads to $M \leq n-1$.

The first proposition below is not strictly necessary for establishing Mermin violations, but it establishes an interesting and nontrivial connection between local Pauli stabilizers and the nonlocal stabilizers $g_j$. The second proposition is the Mermin violation result.

**Proposition 1** *Let $|G\rangle$ be an $n$-qubit symmetric hypergraph state with nonlocal stabilizers $g_j = X_j \prod_{e \in E(j)} C_{e\setminus j}$ for $1 \leq j \leq n$. $|G\rangle$ has a nontrivial local Pauli stabilizing operator if and only if $\prod_j g_j$ is equal to $X^{\otimes n}$, $-X^{\otimes n}$, or $Y^{\otimes n}$.*

**Proposition 2** *Let $|K_n^{m_1,\ldots,m_k}\rangle$ be a symmetric hypergraph state that is $-X^{\otimes n}$-stable and $n-m_j+1$ is even for $1 \leq j \leq k$. The operator $M = \sum_{j=1}^n g_j + \prod_{j=1}^n g_j$ leads to a violation of a Mermin noncontextuality inequality as described above.*

Here is an example to show that the hypothesis that $n-m_j+1$ is even is necessary in the above proposition. The state $\left|K_5^{3,4}\right\rangle$ is $-X^{\otimes n}$-stable, but the assignment of the value $-1$ to classical variables $C_{1,2}, X_3, X_4, X_5$ and the value $+1$ to all other $C_{e\setminus\{j\}}, X_1, X_2$, leads to the value $M = n+1$, so there is no violation. To see that there are infinitely many examples that meet the hypotheses of the proposition, here is a construction for a 2-level completeness family (that is, a family of states of the form $|K_n^{m_1,m_2}\rangle$) of $-X^{\otimes n}$-stable symmetric hypergraph states. Start with any $|K_n^m\rangle$ single level $-X^{\otimes n}$-stable state. Then $\left|K_{n-1}^{m-1}\right\rangle$ and $\left|K_{n-1}^{1,m-1}\right\rangle$ are both $X^{\otimes n}$-stable. From the latter state, the going up sequence

$$\left|K_n^{2,m}\right\rangle, \left|K_{n+1}^{3,m+1}\right\rangle, \left|K_{n+2}^{4,m+2}\right\rangle, \ldots$$

will eventually hit a $-X^{\otimes n}$-stable state, and the parity $n-m_j+1$ is guaranteed to be even for both $j=1,2$. An example is the state $\left|K_{11}^{6,8}\right\rangle$ that arises from the chain that begins with $\left|K_7^4\right\rangle$. Going down leads to $\left|K_6^3\right\rangle$, then appending 1-gates produces $\left|K_6^{1,3}\right\rangle$, then going up produces the sequence $\left|K_7^{2,4}\right\rangle, \left|K_8^{3,5}\right\rangle, \left|K_9^{4,6}\right\rangle, \left|K_{10}^{5,7}\right\rangle$, and finally $\left|K_{11}^{6,8}\right\rangle$.

## 5 Application: error correction

Proposition 3 summarizes a simple construction that uses symmetric hypergraph states with nontrivial local Pauli stabilizer to encode 1-qubit in $n$-qubits in a way that detects and corrects errors of the form $U^{\otimes n}$, sometimes called "collective decoherence". The code presented here may be noteworthy because it does not derive its detection and correction properties from belonging to the decoherence-free subspace [15, 16, 17]. The wider context of error correcting codes for correlated quantum errors is discussed in references [18, 19].

**Proposition 3** *Let $|0_L\rangle$ be an $n$-qubit symmetric hypergraph state with a non-trivial local Pauli stabilizer for some $n \geq 3$, and let $|1_L\rangle = Z_1 Z_2 |0_L\rangle$. The code with logical qubits $|0_L\rangle, |1_L\rangle$ corrects errors*

$$E_0 = \text{Id}^{\otimes n}, E_1 = X^{\otimes n}, E_2 = Y^{\otimes n}, E_3 = Z^{\otimes n}.$$

## 6 Recovery from loss of qubits

In this section we show that symmetric hypergraph states are determined by their reduced density matrices; the necessary minimum cardinality of the subsystems that allow recovery of the original state is governed by the cardinality of the smallest hyperedge. The precise statement is in Proposition 4 below.

Some difficult open problems provide a context for Proposition 4. The closely related problems of quantum marginals $N$-representability are concerned with the general question of when and how a quantum state is determined by its reduced density matrices [20]. The graph state recovery problem asks when an $n$-vertex graph is determined by its "deck" of $(n-1)$-vertex subgraphs [21, 22]. Proposition 4 is a special case (admittedly with many constraints) of the quantum marginals problem, and is a special case of a possible generalization to hypergraphs of the graph recovery problem.

Proposition 4 follows from an analysis of the partial trace operation on hypergraph states in general and on symmetric hypergraph states in particular. Here is the general statement for hypergraph states (stated formally and proved in [10]).

**Partial trace of a hypergraph state.** There are two natural ways to remove a vertex $a$ from a hypergraph $G$. The first way is to erase all the hyperedges that contain $a$, then remove the vertex $a$. The second way is to simply erase the vertex $a$ (thereby shrinking the cardinality of hyperedges that formerly contained that vertex). We call these two hypergraphs $D_a G, S_a G$, where the symbols '$D$','$S$' are mnemonic for "delete" and "shrink" operations, respectively. The partial trace over qubit $a$ of the density matrix $\rho^G = |G\rangle \langle G|$ is the equal mixture of the density matrices for $\rho^{D_a G} = |D_a G\rangle \langle D_a G|, \rho^{S_a G} = |S_a G\rangle \langle S_a G|$. For a symmetric hypergraph state, it does not matter what qubit we trace over, so we may write $DG, SG$, without subscripts, for the delete and shrink operations applied to $G$. The partial trace over any single qubit, say, qubit label 1, of a symmetric hypergraph state is the following.

$$\text{tr}_1(\rho^G) = \frac{1}{2} \left( \rho^{DG} + \rho^{SG} \right) \tag{7}$$

Here is the consequence, in terms of weight sign vectors (3), for the case of symmetric hypergraph states, followed by the statement of the recovery result.

**Lemma 5** *Let $G$ be a symmetric hypergraph on $n$ vertices with corresponding state $|G\rangle$ with weight sign vector $(f_0, f_1, \ldots, f_n)$. Let $f^G$, $f^{DG}$, $f^{SG}$ denote the weight sign vectors for $G$, $DG$, and $SG$, respectively (because $G$ is symmetric,* *it does not matter which vertex we trace over). Weight sign vector entries are the following.*

$$f^G = (f_0, f_1, \ldots, f_n)$$
$$f^{DG} = (f_0, f_1, \ldots, f_{n-1}) \qquad (8)$$
$$f^{SG} = (f_1, f_2, \ldots, f_n) \qquad (9)$$

**Proposition 4** *Let $G, G'$ be symmetric hypergraphs with minimum hyperedge cardinality $k+1$, and let $|G\rangle, |G'\rangle$ be the corresponding hypergraph states. Suppose that $(n-k)$-qubit reduced density matrices are the same for both states, that is,*

$$\mathrm{tr}_{1,2,\ldots,k}(|G\rangle\langle G|) = \mathrm{tr}_{1,2,\ldots,k}(|G'\rangle\langle G'|)$$

*(because the states are symmetric, we may trace over any $k$ qubits). Then $G = G'$.*

## 7 Conclusion and Outlook

We have characterized symmetric hypergraph states that have nontrivial local Pauli symmetry, and have presented applications of these states to nonlocality and error correction.

A natural direction for further pursuit is the larger problem to which the results of this paper constitute a partial solution. Namely, what are *all* possible local unitary stabilizing operators for symmetric hypergraph states, in addition to local Paulis? In our previous work [10], we showed that $|K_4^3\rangle$ has two local unitary symmetries of the form $(aX + bZ)^{\otimes 4}$ that are not local Pauli. Numerical searches have failed to discover any other examples, and in any case, such searches become infeasible for larger numbers of qubits. If $|K_4^3\rangle$ is the only state with local unitary symmetries that are not local Pauli, it would be interesting to know why. While there is a simple interpretation of local unitary symmetries as rotations of the Bloch sphere preserving the Majorana points that represent the state [23], we have so far not been able to connect the Majorana picture with the hypergraph diagram. It would be interesting to achieve this, and also to interpret a physical meaning for the non-Pauli symmetries.

Additional avenues for further investigation are systematic analyses of codes made using hypergraph states and the development of some kind of computational protocol where hypergraph state resources could outperform other known resource states, including graph states, in some meaningful way.

**Acknowledgments.** DL is grateful for helpful discussions with Otfried Gühne and Marcus Huber. This work was supported by National Science Foundation award PHY-1211594 and a Lebanon Valley College Arnold grant and faculty research grants.

## A  Proofs

**Binary expansion convention.** For a positive integer $a$, let $a_j$ denote the coefficient of $2^j$ in the binary expansion of $a$. That is, $a = \sum_{j=0}^{t} a_j 2^j$, where $t$ satisfies $2^t \leq a \leq 2^{t+1} - 1$. Note that $a_j = 0$ for $j > t$.

A main tool is Lucas' Theorem, a little cleaner than the result of Kümmer that we used in our previous paper [10].

**Lucas Theorem (1878 Édouard Lucas [24]).** For nonnegative $m, n$ and prime $p$,
$$\binom{m}{n} = \prod_j \binom{m_j}{n_j} \pmod{p}$$
where $m = \sum_{u=0}^{k} m_j p^j$, $n = \sum_{j=0}^{k} n_j p^j$ are the base $p$ expansions of $m, n$, respectively. A consequence is that $p$ divides $\binom{m}{n}$ if and only if $n_j > m_j$ for some $j$. In particular, this means $\binom{m}{n}$ is even if and only if there is a position where the base 2 expansion of $n$ has a 1 and the base 2 expansion of $m$ has a 0.

**Proof of Lemma 1**

Let $C = \prod_{j=1}^{k} \left( \prod_{e: |e|=m_j} C_e \right)$, so that $|K_n^m\rangle = C |+\rangle^{\otimes n}$. For computational basis vector $|I\rangle$ with $w = \text{wt}(I)$, we have
$$C |I\rangle = (-1)^{\binom{w}{m_1}+\binom{w}{m_2}+\cdots+\binom{w}{m_k}} |I\rangle.$$
so that we have
$$f_w = (-1)^{\binom{w}{m_1}+\binom{w}{m_2}+\cdots+\binom{w}{m_k}}$$
$$e_w = \binom{w}{m_1} + \binom{w}{m_2} + \cdots + \binom{w}{m_k} \pmod{2}$$

From the definitions, then, we have $e = Ag \pmod 2$.

To see that $A = A^{-1}$, consider the $i, j$ entry of $A^2$.
$$(A^2)_{ij} = \sum_{k=0}^{n} \binom{i}{k} \binom{k}{j}$$

If $i = j$, the only nonzero term on the right hand side is $\binom{i}{i}\binom{i}{i} = 1$. If $i < j$, then every term on the right is zero because the factor $\binom{k}{j} = 0$ for $k < j$ and the factor $\binom{i}{k} = 0$ for $k \geq j$.

If $j < i$, we have $A^2 = \sum_{k=j}^{i} \binom{i}{k}\binom{k}{j}$. By Lucas Theorem, every $k$ value for which $\binom{i}{k}\binom{k}{j} \neq 0 \pmod 2$ must satisfy
$$j_\ell = 1 \Rightarrow k_\ell = 1 \Rightarrow i_\ell = 1$$

Choose the largest such $k$. Let $p_1, p_2, \ldots, p_r$ be positions where $k_{p_\ell} = 1$ and $j_{p_\ell} = 0$. The set of all $k$'s that yield nonzero terms is precisely
$$\{j + \sum_i a_i 2^{p_i} : a_i = 0, 1, 1 \leq i \leq r\}.$$

Thus there are precisely $2^r$ such $k$ values, so $(A^2)_{ij} = 0 \pmod 2$. Thus we have proved $A^{-1} = A \pmod 2$.

It immediately follows that $g = Ae \pmod 2$, and the proof is complete.

**Proof of Lemma 2**

[Note: We stated a version of this result for the case where the hyperedge cardinality vector $m$ has only one entry in [10], to use as an observation in the proof of a theorem. Regrettably, the statement in that earlier paper for $Y^{\otimes n}$-stability contained an error. We claimed that $Y^{\otimes n}$-stability implies $n$ must be a multiple of 4. This is false: for example, $\left|K_n^2\right\rangle$ is $Y^{\otimes n}$-stable for all even $n$. Happily, this error does not affect the validity of the proof in which it was made. The statement in this paper corrects the error and generalizes the result to arbitrary hyperedge cardinality vectors.]

Consider the hypergraph $K_n^{m_1,m_2,\ldots,m_k}$, and let $I = i_1 i_2 \ldots i_n$ be a bit string with Hamming weight $w$. For a fixed index $j$ in the range $1 \leq j \leq k$, it is clear from the definition of the generalized controlled-$Z$ operators that the product $\prod_{e:\, |e|=m_j} C_e$ acts on $|I\rangle$ by

$$\left(\prod_{e:\, |e|=m_j} C_e\right) |I\rangle = (-1)^{\binom{w}{m_j}} |I\rangle$$

and hence that

$$\left(\prod_{e \in E} C_e\right) |I\rangle = (-1)^{\left(\binom{w}{m_1}+\binom{w}{m_2}+\cdots+\binom{w}{m_k}\right)} |I\rangle.$$

Statements (i)–(iii) of the Lemma now follow from the simple observations that $X^{\otimes n} |I\rangle = |I^c\rangle$ and $Y^{\otimes n} |I\rangle = (-1)^w i^n |I^c\rangle$, where $I^c$ denotes the bit-wise complement of $I$. For $n$ even, the latter expression reads $Y^{\otimes n} |I\rangle = (-1)^{w+n/2} |I^c\rangle$. It is clear that if $Y^{\otimes n}$ stabilizes the state vector of a symmetric hypergraph state, then $n$ must be even (just look at $Y^{\otimes n} |0\rangle^{\otimes n} = i^n |1\rangle^{\otimes n}$ and note that all state vector coefficients (in the computational basis expansion) for hypergraph states are real).

**Corollary to the proof of Lemma 2.**

Using the same reasoning as above, it is a simple matter to derive palindrome conditions for symmetric hypergraph states that are $-Y^{\otimes n}$-stable or $\pm i Y^{\otimes n}$-stable. It will turn out that such states do not exist, but we need the following palindrome conditions to establish their nonexistence in the proof of Theorem 2.

**More palindrome conditions.** If a symmetric hypergraph state is $-Y^{\otimes n}$-stable then $n$ is even and

$$\sum_j \binom{w}{m_j} = \sum_j \binom{n-w}{m_j} + \frac{n}{2} + w + 1 \tag{10}$$

for $0 \leq w \leq n$. If a symmetric hypergraph state is $iY^{\otimes n}$-stable then $n$ is odd and

$$\sum_j \binom{w}{m_j} = \sum_j \binom{n-w}{m_j} + \frac{n+1}{2} + w \tag{11}$$

for $0 \leq w \leq n$. If a symmetric hypergraph state is $-iY^{\otimes n}$-stable then $n$ is odd and

$$\sum_j \binom{w}{m_j} = \sum_j \binom{n-w}{m_j} + \frac{n+1}{2} + w + 1 \tag{12}$$

for $0 \leq w \leq n$.

**Proof of Theorem 1**

First, observe that it suffices to prove the theorem for the case where the smallest hyperedge cardinality $m_1$ is at least 2, as follows. Let $|G\rangle = |K_n^{m_1,m_2,\ldots,m_k}\rangle$ with $m_1 = 1$, and let $|G'\rangle = Z^{\otimes n} |G\rangle = |K_n^{m_2,m_3,\ldots,m_k}\rangle$. If $U^{\otimes n}$ stabilizes $|G'\rangle$, then $(UZ)^{\otimes n}$ stabilizes $|G\rangle$. We will assume $m_1 \geq 2$ for the remainder of the proof.

Let $|G\rangle$ be a symmetric hypergraph state that meets the hypotheses of the theorem, with smallest hyperedge cardinality at least 2, with weight sign vector $(f_0, f_1, \ldots, f_n)$, and suppose $\bigotimes_j U_j$ stabilizes $|G\rangle$. By permutational invariance, the operator $U_2 \otimes U_1 \otimes \left(\bigotimes_{j \geq 3} U_j\right)$ also stabilizes $|G\rangle$, and therefore

$$\left(U_1 \otimes U_2 \otimes \left(\bigotimes_{j \geq 3} U^j\right)\right) \left(U_2 \otimes U_1 \otimes \left(\bigotimes_{j \geq 3} U^j\right)\right)^\dagger = U_1 U_2^\dagger \otimes U_2 \otimes U_1^\dagger \otimes \mathrm{Id}_{2^{n-2}}$$

stabilizes $|G\rangle$. Let $V = U_1 U_2^\dagger$. We will show that $V \otimes V^\dagger$ must be the identity; it then follows that V must be a phase times the identity. Thus there are phases $e^{it_{ij}}$ such that $U_i = e^{it_{i,j}} U_j$ for all $i, j$, so $\bigotimes_k U_k$ may be written $e^{it} U^{\otimes n}$ for some $t$, where $U$ is (a phase times) any one of the original $U_k$. The remainder of the proof is to show that $V$ must be a phase times the identity.

We begin by rewriting $|G\rangle$, factoring off the first two qubits. We have

$$2^n |G\rangle = \sum_{k=0}^{n-2} (f_k |00\rangle + f_{k+1}(|01\rangle + |10\rangle) + f_{k+2} |11\rangle) \sum_J |J\rangle$$

where $J$ ranges over all $(n-2)$-bit strings. Thus we observe that $V \otimes V^\dagger \otimes \mathrm{Id}_{2^{n-2}} |G\rangle = |G\rangle$ is equivalent to the condition that the vectors

$$(f_j, f_{j+1}, f_{j+1}, f_{j+2}) = f_j |00\rangle + f_{j+1}(|01\rangle + |10\rangle) + f_{j+2} |11\rangle \quad (13)$$

must be $+1$ eigenvectors of $V \otimes V^\dagger$ for $0 \leq j \leq n - 2$. Lemma 6 shows that for most states, the span of these $n - 2$ vectors has dimension at least 3. Since the eigenvalues of $V \otimes V^\dagger$ are (with multiplicities)

$$1, 1, e^{it}, e^{-it}$$

(for some real $t$) we conclude that $V \otimes V^\dagger$ has eigenvalue 1 with, in fact, multiplicity 4, and therefore must be the $4 \times 4$ identity matrix. To complete the proof, we must consider the states that are exceptions to Lemma 6, and show by other means that $V$ is a phase times identity. These exceptional states are $|K_n^n\rangle$ and $|K_n^{2,3,\ldots,n}\rangle$.

Begin by writing $V = cW$ where $c$ is a phase and $W = \begin{bmatrix} a & -b^* \\ b & a^* \end{bmatrix}$ is a special unitary matrix (some $a, b \in \mathbb{C}$ with $|a|^2 + |b|^2 = 1$).

For the state $|K_n^n\rangle$, the weight sign vector is $(1, 1, \ldots, 1, -1)$. There are two $+1$ eigenvectors (13), namely $v_0 = (f_0, f_1, f_1, f_2) = (1, 1, 1, 1)$ and $v_1 = (f_{n-2}, f_{n-1}, f_{n-1}, f_n) = (1, 1, 1, -1)$. It is an elementary computation to verify that

$$\begin{aligned} 0 &= 1 + (-1) \\ &= \langle 00|v_0\rangle + \overline{\langle 11|v_1\rangle} \\ &= \langle 00|V \otimes V^\dagger|v_0\rangle + \overline{\langle 11|V \otimes V^\dagger|v_1\rangle} \\ &= -2(b^*)^2 \end{aligned}$$

so that we conclude $b = 0$. Next,

$$1 = \langle 01|v_1\rangle = \langle 01|V \otimes V^\dagger|v_0\rangle = a^2$$

yields $a^2 = 1$, so $V \otimes V^\dagger$ is the identity.

For the state $|K_n^{2,3,\ldots,n}\rangle$, the $f$ vector is $(1, 1, -1, 1, -1, 1, \ldots)$. Let $v_0 = (f_0, f_1, f_1, f_2) = (1, 1, 1, -1)$ and let $v_1 = (f_1, f_2, f_2, f_3) = (1, -1, -1, 1)$. A similar computation as before yields

$$0 = \langle 11|\left(V \otimes V^\dagger(|v_0\rangle + |v_1\rangle)\right) = -2b^2$$

so $b = 0$. Then we have $1 = \langle 01|V \otimes V^\dagger|v_0\rangle = a^2$, so $a^2 = 1$.

This concludes the proof of Theorem 1.

**Lemma 6 (for Theorem 1)** *Let $|G\rangle =$ be an $n$-qubit symmetric hypergraph state with $n \geq 3$, weight sign vector $(f_0, f_1, \ldots, f_n)$ and with smallest hyperedge cardinality at least 2. With the exception of the states $|K_n^n\rangle$, $|K_n^{2,3,\ldots,n}\rangle$, and $|K_n^2\rangle$, the span of the set of vectors*

$$\{(f_j, f_{j+1}, f_{j+1}, f_{j+2})\colon 0 \leq j \leq n-2\}$$

*has dimension at least 3.*

**Proof of Lemma 6.** It is clear that it suffices to find three indices $j_1, j_2, j_3$ such that the three vectors

$$(f_{j_i}, f_{j_i+1}, f_{j_i+2}) \tag{14}$$

($i = 1, 2, 3$) are independent. We proceed by cases.

Case 0. $m_1 \geq 3$, $n \geq m_1 + 1$. Then the initial entries of the weight sign vector are $(1, 1, 1, -1, X)$ for some $X = \pm 1$. Three vectors of the form (14) for $j_i = 1, 2, 3$ are $(1, 1, 1), (1, 1, -1), (1, -1, X)$. The determinant of the matrix with those vectors in the columns is $-4$ (independent of the value of $X$), so the vectors span 3 dimensions.

Case 1. $m_1 \geq 3$, $n = m_1$. This is the state $|K_n^n\rangle$ and is dealt with in the proof of Theorem 1.

Case 2. $m_1 = 2$. Then the initial entries of the weight sign vector are $(1, 1, -1, X, Y)$, some $X = \pm 1$, $Y = \pm 1$. Three vectors of the form (14) for $j_i = 1, 2, 3$ are $(1, 1, -1), (1, -1, X), (-1, X, Y)$. The determinant of the matrix with those columns is $-2(X + Y)$. The degenerate cases are the following.

Case 2a. $X = 1, Y = -1$, weight sign vector begins $(1, 1, -1, 1, -1)$. One readily checks that if the weight sign vector does not continue to alternate (two $+1$ entries followed by two $-1$ entries, etc) then there are 3 independent vectors $(f_j, f_{j+1}, f_{j+2})$, so we assume the alternating pattern continues. This is the state $|K_n^{2,3,\ldots,n}\rangle$. This is dealt with in the proof of the theorem.

Case 2b. $X = -1, Y = 1$, weight sign vector begins $(1, 1, -1, -1, 1)$. Again, one verifies that the degenerate span of the vectors $(f_j, f_{j+1}, f_{j+2})$ (that is, a span of dimension 2) implies that the weight sign vector continues as a sequence of pairs of $+1$s and pairs of $-1$s. This is the graph state $|K_n^2\rangle$.

Summary: we have shown that there are some states that fail to have three independent vectors of the form $(f_j, f_{j+1}, f_{j+2})$.

- $|K_n^n\rangle$
- $|K_n^{2,3,\ldots,n}\rangle$
- $|K_n^2\rangle$ (graph states)

This concludes the proof.

**Proof of Theorem 2**

By Theorem 1, any nontrivial local Pauli stabilizer must be of the form $cX^{\otimes n}$, $cY^{\otimes n}$, or $cZ^{\otimes n}$ for $c = \pm 1, \pm i$. Here is a summary of how we rule out all the possibilities except for $X^{\otimes n}$, $-X^{\otimes n}$, $Y^{\otimes n}$.

We rule out $\pm iX^{\otimes n}$ because state vector coefficients are real.

We rule out all $cZ^{\otimes n}$ because $Z$ acts diagonally and $Z^{\otimes n}$ acts on computational basis vector $|I\rangle$ by $(-1)^{\text{wt}(I)}$, so any odd weight standard basis vector gets its sign changed.

We rule out $-Y^{\otimes n}$ by putting $w = n/2$ in the palindrome condition (10) given in the proof of Lemma 2 above. The equation implies $1 = 0$, which is a contradiction.

Finally, we rule out $\pm iY^{\otimes n}$ by putting $w = 0, n$ in the palindrome condition (11) or (12) (for coefficient $i, -i$, respectively) given in the proof of Lemma 2 above. The two equations imply $n$ is even, which is a contradiction.

**Proof of Lemma 3**

The basic observation at the heart of this proof is that the total parity of any triple $\binom{a-1}{b-1}, \binom{a-1}{b}, \binom{a}{b}$ of entries in Pascal's triangle is even: reading the Pascal recursion identity $\binom{a-1}{b-1} + \binom{a-1}{b} = \binom{a}{b}$ mod 2 says that any of the three entries is equal to the sum of the other two.

Now suppose $|K_n^m\rangle$ is $X^{\otimes n}$-stable. We show that palindrome condition (4) holds for $|K_{n-1}^{m-1}\rangle$. For $0 \leq w \leq n-1$, we have the following equations mod 2.

$$\sum_j \binom{w-1}{m_j-1} = \sum_j \left(\binom{w-1}{m_j} + \binom{w}{m_j}\right) \quad \text{(basic observation)}$$
$$= \sum_j \left(\binom{n-(w-1)}{m_j} + \binom{n-w}{m_j}\right) \quad \text{(by (4) for } |K_n^m\rangle\text{)}$$
$$= \sum_j \binom{n-w}{m_j-1} \quad \text{(basic observation)}$$
$$= \sum_j \binom{(n-1)-(w-1)}{m_j-1}.$$

This establishes part (a) of the Lemma.

Next, suppose $|K_n^m\rangle$ is $-X^{\otimes n}$-stable. The same derivation above shows that palindrome condition (4) holds for $|K_{n-1}^{m-1}\rangle$, with the minor alteration of apply-ing palindrome condition (5) on the second line in place of (4).

$$\sum_j \binom{w-1}{m_j-1} = \sum_j \left(\binom{w-1}{m_j} + \binom{w}{m_j}\right) \quad \text{(basic observation)}$$

$$= \sum_j \left(\binom{n-(w-1)}{m_j} + 1 + \binom{n-w}{m_j} + 1\right) \quad \text{(by (5) for } |K_n^m\rangle)$$

$$= \sum_j \binom{n-w}{m_j-1} \quad \text{(basic observation)}$$

$$= \sum_j \binom{(n-1)-(w-1)}{m_j-1}.$$

This establishes part (b) of the Lemma.

Finally, suppose $|K_n^m\rangle$ is $Y^{\otimes n}$-stable. Once again, a small modification of the same derivation used above shows that palindrome condition (5) holds for $|K_{n-1}^{m-1}\rangle$. For $0 \le w \le n-1$, we have the following equations mod 2.

$$\sum_j \binom{w-1}{m_j-1} = \sum_j \left(\binom{w-1}{m_j} + \binom{w}{m_j}\right) \quad \text{(basic observation)}$$

$$= \sum_j \left(\binom{n-(w-1)}{m_j} + w - 1 + \binom{n-w}{m_j} + w\right) \quad \text{(by (6) for } |K_n^m\rangle)$$

$$= \sum_j \binom{n-w}{m_j-1} + 1 \quad \text{(basic observation)}$$

$$= \sum_j \binom{(n-1)-(w-1)}{m_j-1} + 1.$$

This establishes part (c) of the Lemma, and completes the proof.

**Proof of Lemma 4**

Suppose that $|K_n^m\rangle$ is $X^{\otimes n}$-stable. We show that either palindrome condition (4) or (5) holds for $|K_{n+1}^{m+1}\rangle$. We continue to make use of the basic observation from the proof of Lemma 3 about total even parity in certain triples of entries in Pascal's triangle.

By palindrome condition (4) for $|K_n^m\rangle$, we have, for $0 \le w \le n$

$$\sum_j \binom{w}{m_j} = \sum_j \binom{n-w}{m_j} \pmod{2}.$$

Applying the basic observation to the left and right hand sides, then rearranging terms, we have the following mod 2 equations for $0 \le w \le n$.

$$\sum_j \left(\binom{w}{m_j+1} + \binom{w+1}{m_j+1}\right) = \sum_j \left(\binom{(n+1)-w}{m_j+1} + \binom{(n+1)-(w+1)}{m_j+1}\right)$$

$$\underbrace{\sum_j \binom{w}{m_j+1}}_{A} + \underbrace{\sum_j \binom{w+1}{m_j+1}}_{B} = \underbrace{\sum_j \binom{(n+1)-w}{m_j+1}}_{A'} + \underbrace{\sum_j \binom{(n+1)-(w+1)}{m_j+1}}_{B'}$$

It follows that exactly one of the following two conditions in square brackets must hold mod 2 for $0 \leq w \leq n$.

$$[A = A' \text{ AND } B = B'] \text{ OR } [A = A' + 1 \text{ AND } B = B' + 1]$$

One can readily see that if the condition in square brackets on the left holds, then palindrome condition (4) holds for $\left|K_{n+1}^{m+1}\right\rangle$, and that if the condition in square brackets on the right holds, then palindrome condition (5) holds for $\left|K_{n+1}^{m+1}\right\rangle$. We conclude that $\left|K_{n+1}^{m+1}\right\rangle$ is either $X^{\otimes n}$-stable or is $-X^{\otimes n}$-stable.

Now suppose that $|K_n^m\rangle$ is $-X^{\otimes n}$-stable. We show that palindrome condition (6) holds for $\left|K_{n+1}^{m+1}\right\rangle$. The argument is the same as above, with the modification that we start with palindrome condition (5) for $|K_n^m\rangle$. We have, for $0 \leq w \leq n$

$$\sum_j \binom{w}{m_j} = \sum_j \binom{n-w}{m_j} + 1 \pmod 2.$$

Applying the basic observation to the left and right hand sides, then rearranging terms, we have the following mod 2 equations for $0 \leq w \leq n$.

$$\sum_j \left(\binom{w}{m_j+1} + \binom{w+1}{m_j+1}\right) + 1 = \sum_j \left(\binom{(n+1)-w}{m_j+1} + \binom{(n+1)-(w+1)}{m_j+1}\right) + 1$$

$$\underbrace{\sum_j \binom{w}{m_j+1}}_{A} + \underbrace{\sum_j \binom{w+1}{m_j+1}}_{B} = \underbrace{\sum_j \binom{(n+1)-w}{m_j+1}}_{A'} + \underbrace{\sum_j \binom{(n+1)-(w+1)}{m_j+1}}_{B'} + 1$$

It follows that exactly one of the following two conditions in square brackets must hold mod 2 for $0 \leq w \leq n$.

$$[A = A' \text{ AND } B = B' + 1] \text{ OR } [A = A' + 1 \text{ AND } B = B']$$

This implies that one of the two sequences of equations

$$f_w = \begin{cases} f_{n-w} & w \text{ odd} \\ f_{n-w} + 1 & w \text{ even} \end{cases}$$

or

$$f_w = \begin{cases} f_{n-w} + 1 & w \text{ odd} \\ f_{n-w} & w \text{ even} \end{cases}$$

must hold for the weight sign vector for all indices $w$. This is the same as saying that palindrome condition (6) holds. We conclude that $\left|K_{n+1}^{m+1}\right\rangle$ is $Y^{\otimes n}$-stable.

**Proof of Theorem 3**

[Note: the 'if' direction of statements (i) and (iii) are proved in our previous paper [10]. Thus, it is only strictly necessary to prove the remaining parts of the theorem. However, in order to provide a self-contained exposition, and also because we now have cleaner, shorter, arguments, we present here a complete proof of the entire theorem.]

First, we prove the 'if' direction of statement (ii). Let $m = 2^t$ for some $t \geq 0$ and let $n = (2s+1)2^t + 2^t - 1$ for some $s \geq 0$. The hypotheses imply the binary expansions of $m, n$ satisfy the following.

$$n_t = 1 \qquad\qquad n_{t-1} = n_{t-2} = \cdots = n_0 = 1$$
$$m_t = 1 \qquad\qquad m_{t-1} = m_{t-2} = \cdots = m_0 = 0$$

Observe that, for any $w$ in the range $0 \leq w \leq n$, we have

$$w_t = 0 \Leftrightarrow (n-w)_t = 1.$$

By Lucas' Theorem, we have

$$\binom{w}{m} = 0 \Leftrightarrow w_t = 0, \qquad \binom{n-w}{m} = 1 \Leftrightarrow (n-w)_t = 1.$$

Thus, the palindrome condition 6 is satisfied, so we conclude that $|K_n^m\rangle$ is $-X^{\otimes n}$-stable, as desired.

The 'if' direction of statements (i) and (iii) now follow immediately by the going down and going up Lemmas 3, 4, respectively.

Next, we prove the 'only if' direction of statement (i). Suppose $|K_n^m\rangle$ is $X^{\otimes n}$-stable, with $m \geq 1$ and $n \geq m$. Choose $t \geq 0$ such that

$$2^t \leq m \leq 2^{t+1} - 1.$$

Applying the going down Lemma 3 to the palindrome condition (4) gives

$$\binom{w}{m-\alpha} = \binom{n-\alpha-w}{m-\alpha} \pmod{2}$$

for $0 \leq w \leq n - \alpha$, $0 \leq \alpha \leq m - 1$. Putting $w = 0$ yields

$$0 = \binom{n-\alpha}{m-\alpha} \pmod{2}$$

for $0 \leq \alpha \leq m$. Putting $\alpha = m - 2^s$ for $0 \leq s \leq t$, and applying Lucas' theorem, implies that the binary expansion of $n - \alpha = n - m + 2^s$ must have a 0 in position $s$, so the binary expansion of $n - m$ has a 1 in position $s$, for $0 \leq s \leq t$. Thus we have

$$n - m = k2^{t+1} + 2^{t+1} - 1$$

for some $k \geq 0$, which is the same as

$$n = s2^{t+1} + m - 1$$

for some $s \geq 1$, as desired. Thus both the 'if' and 'only if' directions of statement (i) have been established.

To establish 'only if' direction of statement (ii), suppose that $|K_n^m\rangle$ is $-X^{\otimes n}$-stable. By the going down Lemma 3, we have $|K_{n-1}^{m-1}\rangle$ is $X^{\otimes n}$-stable. By the 'only if' direction of statement (i), there exist $t \geq 0$, $k \geq 1$ such that $2^t \leq m - 1 \leq 2^{t+1} - 1$ and $n - 1 = k2^{t+1} + (m-1) - 1$. If $m \neq 2^{t+1}$, then $|K_n^m\rangle$ would be $X^{\otimes n}$-stable. But the only vector that is both $X^{\otimes n}$-stable and also $-X^{\otimes n}$-stable is the zero vector, so we must have $m = 2^{t+1}$. If $k$ is even, then we have $n = k'2^{t+2} + 2^{t+1} - 1$ for some $k' \geq 1$, so once again we would have $|K_n^m\rangle$ is $X^{\otimes n}$-stable. Since this is impossible, we conclude that $k$ is odd, as desired.

Finally, we observe that the 'only if' direction of statement (iii) holds by applying the going up Lemma 4 to statement (ii). This concludes the proof of the theorem.

**Proof of Theorem 4**

This theorem follows immediately from the going down and going up Lemmas 3, 4, together with the palindrome conditions Lemma 2.

**Proof of Proposition 1**

First we generalize Lemma 2 of Gühne et al. [6]. (In [6], the hypothesis on set $K$ below is $K \subseteq e$.)

**Lemma 7** *Lemma Let $K$ be a subset of $\{1, 2, \ldots, n\}$ and let $e$ be a hyperedge. Then we have*

$$C_e \prod_{k \in K} X_k = \prod_{k \in K} X_k \prod_{f \subseteq e \cap K} C_{e \setminus f}.$$

**Proof.** We have

$$C_e \prod_{k \in K} X_k = C_e \prod_{k \in e \cap K} X_k \prod_{k \in K \setminus e} X_k \tag{15}$$

$$= \prod_{k \in e \cap K} X_k \prod_{f \subseteq e \cap K} C_{e \setminus f} \prod_{k \in K \setminus e} X_k \quad \text{(by [6] Lemma 2)} \tag{16}$$

$$= \prod_{k \in K} X_k \prod_{f \subseteq e \cap K} C_{e \setminus f} \tag{17}$$

because the $X_k$'s with $k \in K \setminus e$ commute with all the $C_{e \setminus f}$'s. This completes the proof of Lemma 7.

Next, we derive a general formula for $\prod g_j$, where $g_j = \left(\prod_{e \in E} C_e\right) X_j \left(\prod_{e \in E} C_e\right)$.

**Lemma 8** *We have*

$$\prod_{j=1}^n g_j = (-1)^{|E|} \prod_{j=1}^n X_j \prod_{e \in E} \prod_{\emptyset \neq f \subsetneq e} C_{e \setminus f}. \tag{18}$$

**Proof.** We have

$$\prod_{j=1}^n g_j = \prod_{j=1}^n \left[ \left(\prod_{e \in E} C_e\right) X_j \left(\prod_{e \in E} C_e\right) \right] \quad \text{(definition of } g_j\text{)}$$

$$= \left(\prod_{e \in E} C_e\right) \left(\prod_j X_j\right) \left(\prod_{e \in E} C_e\right)$$

$$= \left(\prod_j X_j\right) \left(\prod_{e \in E} \prod_{f \subseteq e} C_{e \setminus f}\right) \left(\prod_{e \in E} C_e\right) \quad \text{(Lemma 7 applied to } K = \{1, 2, \ldots, n\}\text{)}$$

$$= (-1)^{|E|} \left(\prod_j X_j\right) \left(\prod_{e \in E} \prod_{\emptyset \neq f \subsetneq e} C_{e \setminus f}\right)$$

where the last expression is obtained by factoring out

$$\prod_e \prod_{f=e} C_{e\setminus f} = \prod_e C_\emptyset = (-1)^{|E|}$$

from the expression in the middle set of parentheses and also factoring out

$$\prod_e \prod_{f=\emptyset} C_{e\setminus f} = \prod_e C_e$$

and canceling with the entire expression in the right-most set of parentheses. This completes the proof of Lemma 8.

Now we prove Proposition 1. Clearly, we only need to prove the 'only if' direction. The heart of the proof is to count multiplicities of factors $C_{e_0\setminus f_0}$ in the product

$$\prod_{e\in E} \prod_{\emptyset\neq f\subsetneq e} C_{e\setminus f} \qquad (19)$$

in the expression (18). To be precise, we determine the cardinality of the set

$$\{(e,f)\colon e\in E, \emptyset\neq f\subsetneq e, C_{e\setminus f} = C_{e_0\setminus f_0}\}$$

for a chosen hyperedge $e_0 \in E$ and proper subset $f_0$ ($\emptyset \neq f_0 \subsetneq e_0$). Setting $d = |e_0| - |f_0|$, one readily sees that this number is given by

$$\sum_{j\colon m_j \geq d+1} \binom{n-d}{m_j - d} \qquad (20)$$

and we must consider $d$ in the range $1 \leq d \leq m_k$, where $m_k$ is the largest hyperedge cardinality.

Suppose now that $|K_n^{m_1,\ldots,m_k}\rangle$ is $X^{\otimes n}$-stable or $-X^{\otimes n}$-stable. By the going down Lemma 3, we have

$$\left|K_{n-1}^{m_1-1,\ldots,m_k-1}\right\rangle, \left|K_{n-2}^{m_1-2,\ldots,m_k-2}\right\rangle, \ldots \left|K_{n-m_k-1}^{m_k-(m_k-1)}\right\rangle$$

are all $X^{\otimes n}$-stable. By the palindrome condition (4), the sum (20) is even for $1 \leq d \leq m_k - 1$. We conclude that

$$\prod_{e\in E} \prod_{\emptyset\neq f\subsetneq e} C_{e\setminus f} = \mathrm{Id}$$

and therefore

$$\prod_j g_j = (-1)^{|E|} X^{\otimes n} = \begin{cases} X^{\otimes n} & \text{when } |K_n^m\rangle \text{ is } X^{\otimes n}\text{-stable} \\ -X^{\otimes n} & \text{when } |K_n^m\rangle \text{ is } -X^{\otimes n}\text{-stable} \end{cases}$$

(because $|E| = \sum_j \binom{n}{m_j}$ is even or odd depending on whether $|K_n^m\rangle$ is $X^{\otimes n}$-stable or $-X^{\otimes n}$-stable, respectively, by Lemma 2).

Now suppose that $|K_n^{m_1,\ldots,m_k}\rangle$ is $Y^{\otimes n}$-stable. By the going down Lemma 3, we have

$$\left|K_{n-2}^{m_1-2,\ldots,m_k-2}\right\rangle, \left|K_{n-3}^{m_1-3,\ldots,m_k-3}\right\rangle, \ldots \left|K_{n-m_k-1}^{m_k-(m_k-1)}\right\rangle$$

are all $X^{\otimes n}$-stable. By the palindrome condition (4), the sum (20) is even for $2 \leq d \leq m_k - 1$, so all factors $C_{e \setminus f}$ in the product (18) cancel except possibly those corresponding to $d = 1$. We have

$$\prod_{e \in E} \prod_{\emptyset \neq f \subsetneq e} C_{e \setminus f} = \prod_e \prod_{a \in e} Z_a.$$

Every vertex $j$ is an element of $\sum_{j=1}^{k} \binom{n-1}{m_j - 1}$ hyperedges. Applying palindrome condition (5) to the $-X^{\otimes(n-1)}$-stable state $K_{n-1}^{m-1}$ (guaranteed by the going down Lemma 3), we see that $\sum_{j=1}^{k} \binom{n-1}{m_j - 1}$ is odd. Therefore

$$\prod_{e \in E} \prod_{\emptyset \neq f \subsetneq e} C_{e \setminus f} = Z^{\otimes n}$$

and we have

$$\prod_j g_j = (-1)^{|E|} (XZ)^{\otimes n}$$
$$= (-1)^{|E|} (-i)^n Y^{\otimes n}$$
$$= (-1)^{\sum_j \binom{n}{m_j} + n + n/2} Y^{\otimes n}$$
$$= Y^{\otimes n}$$

by palindrome condition (6). This concludes the proof of Proposition 1.

**Proof of Proposition 2**

Assume on the contrary that there exists an assignment of values $\pm 1$ to classical variables $X_j, C_{e \setminus j}$ such that the value of $M = \sum_j g_j + \prod_j g_j$ is $n + 1$ for some symmetric hypergraph state with $-X^{\otimes n}$ symmetry. By Proposition 1, we have

$$\prod_j g_j = -\prod_j X_j.$$

Let $C_j = \prod_{e \in E(j)} C_{e \setminus j}$, so we have

$$M = \sum_j X_j C_j - \prod_j X_j.$$

From this it follows (because the value of $M$ is $n + 1$) that the sets $\{j \colon X_j = -1\} = \{j \colon C_j = -1\}$ both contain an odd number of elements, and therefore

$$\prod_j C_j = -1. \tag{21}$$

The remainder of the proof is a parity argument that shows this leads to a contradiction. The heart of the matter is the following Claim.

**Claim.** Let $e_0 \in E$ and $a_0 \in e$. The number of pairs $(e, a)$ in the set

$$S(e_0, a_0) = \{(e, a) \colon e \in E, a \in e, C_{e \setminus a} = C_{e_0 \setminus a_0}\}$$

is even.

That the Claim holds follows immediately from the hypothesis that $n - m_j + 1$ is even, $1 \leq j \leq k$, and the observation that for $|e_0| = m_j$ and any $a_0 \in e$, the cardinality of the set $S(e_0, a_0)$ is $\binom{n - m_j + 1}{1} = n - m_j + 1$.

Let $T \subseteq E \times V$ denote the set of pairs $(e, a)$ such that $a \in e$, and define an equivalence relation $\sim$ on the $T$ by $(e, a) \sim (f, b)$ if $e \setminus a = f \setminus b$. Now consider an array $A$ with rows indexed by qubits labels $\{1, 2, \ldots, n\}$ and with columns indexed by equivalence classes $T/\sim$. In the array position in row $j$ and column $(e, a)$, put the value $\pm 1$ of $C_{e \setminus a}$ if $j \notin e \setminus a$, and put the value 1 otherwise. With these array entries so constructed, the value of $C_j$ is the product of entries in row $j$. It follows by (21) above that the product of all the entries in array $A$ is $-1$. But by the Claim above, there are an even number of $-1$ entries in every column of the array $A$, so the product of all the entries in array $A$ is 1. From this contradiction we conclude that the Proposition holds.

**Proof of Proposition 3**

Let $|0_L\rangle$ meet the hypotheses. By Theorem 2, the local Pauli operator that stabilizes $|0_L\rangle$ is one of $X^{\otimes n}$, $-X^{\otimes n}$, $Y^{\otimes n}$.

In straightforward calculations, one verifies that the error correction criterion given in Nielsen-Chuang [25], Theorem 10.1, holds. One checks that for each pair of errors $E_i, E_j$, $0 \leq i, j \leq 3$, that $PE_i^\dagger E_j P$ is a scalar multiple $\alpha_{ij}$ times $P$, where $P = |0_L\rangle\langle 0_L| + |1_L\rangle\langle 1_L|$ is the projector onto the code space, and that the matrix $(\alpha_{ij})$ is Hermitian. Here are the matrices $\alpha_{ij}$ for the three cases.

| Pauli stabilizer | $(\alpha_{ij})$ |
|---|---|
| $X^{\otimes n}$ | $\begin{bmatrix} 1 & 1 & 0 & 0 \\ 1 & 1 & 0 & 0 \\ 0 & 0 & 1 & i^n \\ 0 & 0 & -i^n & 1 \end{bmatrix}$ |
| $-X^{\otimes n}$ | $\begin{bmatrix} 1 & -1 & 0 & 0 \\ -1 & 1 & 0 & 0 \\ 0 & 0 & 1 & -i^n \\ 0 & 0 & i^n & 1 \end{bmatrix}$ |
| $Y^{\otimes n}$ | $\begin{bmatrix} 1 & 0 & 1 & 0 \\ 0 & 1 & 0 & (-i)^n \\ 1 & 0 & 1 & 0 \\ 0 & -(-i)^n & 0 & 1 \end{bmatrix}$ |

The calculations that verify $PE_i^\dagger E_j P = \alpha_{ij} P$ use the properties established in the following Lemma.

**Lemma 9 (for Proposition 3)** *Suppose that a $n$-qubit symmetric hypergraph state $|0_L\rangle$ is $X^{\otimes n}$-stable, $-X^{\otimes n}$-stable, or $Y^{\otimes n}$-stable. Let $|1_L\rangle = Z_1 Z_2 |0_L\rangle$. The following properties hold.*

1. *$|1_L\rangle$ is stabilized by the same local Pauli operator as $|0_L\rangle$*
2. *$\langle 0_L | 1_L \rangle = 0$*
3. *$\langle 0_L | Z^{\otimes n} | 1_L \rangle = 0$*
4. *$0 = \langle 0_L | Z^{\otimes n} | 0_L \rangle = \langle 1_L | Z^{\otimes n} | 1_L \rangle$*

Comments: Parts 1 and 2 hold for any hypergraph state, symmetric or not. Other parts require permutation invariance. Here is a proof of the Lemma for the $X^{\otimes n}$ case. Proofs for the other cases follow by the same logic, with minor changes.

1. This follows immediately from the fact that $Z_1 Z_2$ commutes with $X^{\otimes n}$.

2. Write
$$|0_L\rangle = \sum_J (a_{00J}|00J\rangle + a_{01J}|01J\rangle + a_{10J}|10J\rangle + a_{11J}|11J\rangle)$$
where $J$ ranges over $(n-2)$-bit strings, so that we have
$$|1_L\rangle = \sum_J (a_{00J}|00J\rangle - a_{01J}|01J\rangle - a_{10J}|10J\rangle + a_{11J}|11J\rangle).$$
Since $\langle rsJ|r's'J'\rangle = \delta_{rr'}\delta_{ss'}\delta_{JJ'}$ (that is, cross terms in the inner product cancel), and all the coefficients have value $a_{rsJ} = \pm 1$, we have
$$\langle 0_L|1_L\rangle = \sum_J (a_{00J}^2 - a_{01J}^2 - a_{10J}^2 + a_{11J}^2)$$
$$= \sum_J (1 - 1 - 1 + 1) = 0.$$

3. Using the above expressions for $|0_L\rangle, |1_L\rangle$, we have
$$Z^{\otimes n}|1_L\rangle = II \otimes Z^{\otimes(n-2)}|0_L\rangle$$
$$= \sum_J (-1)^{\text{wt}(J)} (a_{00J}|00J\rangle - a_{01J}|01J\rangle - a_{10J}|10J\rangle + a_{11J}|11J\rangle).$$
As in the previous part, cross terms in the inner product cancel, so we have
$$\langle 0_L|Z^{\otimes n}|1_L\rangle = \sum_J (-1)^{\text{wt}(J)} (a_{00J}^2 + a_{01J}^2 + a_{10J}^2 + a_{11J}^2)$$
$$= 4 \sum_J (-1)^{\text{wt}(J)}$$
$$= 4 \sum_{w=0}^{n-2} (-1)^w \binom{n-2}{w}.$$
The latter sum is zero because it is the alternating sum of terms in row $n-2$ of Pascal's triangle.

4. In terms of the computational basis expansion
$$|0_L\rangle = \sum_J a_I|I\rangle$$
(where $I$ ranges over $n$-bit strings), we have, by a similar derivation as in the previous part,
$$\langle 0_L|0_L\rangle = 4\sum_I (-1)^{\text{wt}(I)} = 4\sum_{w=0}^n (-1)^n \binom{n}{w} = 0.$$
The second equality in the statement of part 4 follows from
$$\langle 1_L|1_L\rangle = \langle 0_L|(Z_1Z_2)(Z_1Z_2)|0_L\rangle = \langle 0_L|0_L\rangle.$$

This completes the proof of the Lemma and also the proof of the Proposition.

**Proof of Lemma 5**

Let $G$ be a symmetric hypergraph on $n$ vertices and let $DG, SG$ be the $(n-1)$-vertex hypergraphs made by "delete" and "shrink" operations, respectively, on $G$. We will write

$$e^G, f^G, g^G, \quad e^{DG}, f^{DG}, g^{DG}, \quad e^{SG}, f^{SG}, g^{SG}$$

for the vectors $e, f, g$ (defined in Lemma 1) associated with states $|G\rangle, |DG\rangle, |SG\rangle$, respectively.

Because $DG$ is formed by removing a qubit and all the hyperedges that contain that qubit, $DG$ is complete in all the same levels in which $G$ is complete except for the top level $n$. Thus we have $g^{DG} = (g_0, g_1, \ldots, g_{n-1})$. By Lemma 1, we have $e^{DG} = A_{n-1} g^{DG} = (e_0, e_1, \ldots, e_{n-1})$ (mod 2), where $A_{n-1} = \{\binom{i}{j}\}_{0 \leq i,j \leq n-1}$. Thus we have $f_w^{DG} = f_w^G$ for $0 \leq w \leq n-1$, so (8) is established.

Because $SG$ is formed by deleting a qubit from the vertex set of $G$ and all the hyperedges of $G$ that contain that qubit, any hyperedge of cardinality $m$ in $G$ that contains the qubit that is to be deleted becomes a hyperedge of cardinality $m-1$ in $SG$. It follows that $g^{SG} = (g_0 + g_1, g_1 + g_2, \ldots, g_{n-1} + g_n)$. Thus

$$\begin{aligned} e^{SG} &= A_{n-1} g^{SG} \pmod{2} \\ &= A_{n-1}(g_0, g_1, \ldots, g_{n-1}) + A_{n-1}(g_1, g_2, \ldots, g_n). \end{aligned}$$

It follows that for $0 \leq w \leq n-1$, we have (mod 2)

$$\begin{aligned} e_w^{SG} &= \sum_{k \,:\, g_k = 1} \left( \binom{w}{k-1} + \binom{w}{k} \right) \\ &= \sum_{k \,:\, g_k = 1} \binom{w+1}{k} \\ &= e_{w+1}^G. \end{aligned}$$

Thus we have $f_w^{SG} = f_{w+1}^G$ for $0 \leq w \leq n-1$, so (9) is established. This completes the proof of the Lemma.

**Proof of Proposition 4**

Here we give a complete proof for the case $k = 1$, then outline the proof for the general case.

The assumption that the smallest hyperedge cardinality for $G$ is at least 2 means that the weight sign vector for $G$ has $f_0 = f_1 = 1$ (throughout this proof, we omit the superscript $G$ and write $f$ for $f^G$). By Lemma 5, it follows that the first column of $\rho^{DG}$ is $(f_0, f_1, \ldots, f_{n-1})^T$ and the first column of $\rho^{SG}$ is $(f_1, f_2, \ldots, f_n)^T$, where $T$ denotes transpose. By (7), it follows that the first column of $\text{tr}_a(\rho^G)$ is $\frac{1}{2}(f_0 + f_1, f_1 + f_2, \ldots, f_{n-1} + f_n)^T$. Let us write $v_j$ to denote the entries of this column vector $(v_0, v_1, \ldots, v_{n-1})^T$, so that we have $2v_j = f_j + f_{j+1}$ for $0 \leq j \leq n-1$. Beginning with the base case values $f_0 = f_1 = 1$, we proceed inductively: having solved for the values $f_0, \ldots, f_j$, we have $f_{j+1} = 2v_j - f_j$. In this way we recover the weight sign vector for $|G\rangle$ from $\text{tr}_a(\rho^G)$. This establishes the proposition for the case $k = 1$.

For the general case, we assume that the smallest hyperedge cardinality for $G$ is at least $k+1$. We can view $D, S$ as operators on density matrices on symmetric hypergraph states, and write equation (7)) as

$$\text{tr}_1 = \frac{1}{2}(D + S).$$

It is easy to see that $D, S$ commute, so that the partial partial trace operation over a subsystem of $k$ qubits is given by

$$\text{tr}_{1,2,\ldots,k} = \frac{1}{2^k}(D+S)^k = \sum_{j=0}^{k} \binom{k}{j} D^j S^{k-j}.$$

Applying Lemma 5 inductively, we obtain the weight sign vector for $\left|D^j S^{k-j} G\right\rangle$.

$$f^{D^j S^{k-j}} = (f_{k-j}, f_{k-j+1}, \ldots, f_{n-j})$$

From this, one readily calculates values $v_j$ in the first column of $\text{tr}_{1,\ldots,k}(\rho^G)$ for $0 \leq j \leq n-k$.

$$v_j = \sum_{\ell=0}^{k} f_{j+\ell} \binom{k+1}{\ell}$$

Recovering the weight sign vector for $|G\rangle$ from the $(n-k)$-qubit reduced density matrix $\text{tr}_{1,\ldots,k}(\rho^G)$ is in principle the same as for the simple case $k=1$. Starting with the base case values $f_0 = f_1 = \cdots = f_k = 1$, we proceed inductively: having solved for the values $f_0, \ldots, f_{k+j-1}$, we have $f_{k+j} = 2^k v_j - \sum_{\ell=0}^{k-1} f_{j+\ell} \binom{k+1}{\ell}$. This concludes the sketch of the proof of the general case.